\documentclass[aps, prb, twocolumn, amsmath,amssymb]{revtex4}
\usepackage{graphicx}
\usepackage{bm}

\begin{document}
\bibliographystyle{apsrev}

\title{Smeared phase transition in a three-dimensional Ising model with planar defects: Monte-Carlo simulations}

\author{Rastko Sknepnek}
\affiliation{Department of Physics, University of Missouri - Rolla, Rolla, MO, 65409}

\author{Thomas Vojta}
\affiliation{Department of Physics, University of Missouri - Rolla, Rolla, MO, 65409}

\begin{abstract}
We present results of large-scale Monte Carlo simulations for a three-dimensional Ising model with
short range interactions and planar defects, i.e., disorder perfectly correlated in two
dimensions. We show that the phase transition in this system is smeared, i.e., there is no single
critical temperature, but different parts of the system order at different temperatures. This is
caused by effects similar to but stronger than Griffiths phenomena. In an infinite-size sample
there is an exponentially small but finite probability to find an arbitrary large region devoid of
impurities. Such a rare region can develop true long-range order while the bulk system is still in
the disordered phase. We compute the thermodynamic magnetization and its finite-size effects, the
local magnetization, and the probability distribution of the ordering temperatures for different
samples. Our Monte-Carlo results are in good agreement with a recent theory based on extremal
statistics.
\end{abstract}

\maketitle

\section{Introduction}
\label{sec:Int}

The influence of disorder on a phase transition is an important and still partially open problem.
Historically, the first attempts to address this question resulted in the belief that any kind of
disorder would destroy a critical point because the system would divide itself into regions which
independently undergo the phase transition at different temperatures. Therefore, there would not
be a unique critical temperature for the system, but the phase transition would be smeared over an
interval of temperatures. The singularities of thermodynamic quantities, which are the typical
sign of a phase transition, would also be smeared (see Ref.\ \onlinecite{Grinstein} and references
therein).

However, it soon became clear that this belief was mistaken: in systems with weak short-range
correlated disorder the phase transition remains sharp. Harris proposed a simple, heuristic
criterion\cite{Harris:74} for the influence of disorder on a critical point: if $\nu\ge2/d$, where
$\nu$ is the correlation length critical exponent and $d$ the spatial dimensionality, the disorder
does not affect the critical behavior. In this case, the randomness decreases under coarse
graining, and the system effectively looks homogeneous on large length scales. The critical
behavior is identical to that of the clean system, i.e., the clean renormalization group fixed
point is stable against disorder. The relative widths of the probability distributions of the
macroscopic observables tend to zero in thermodynamic limit, i.e., they are self-averaging.

Even if the Harris criterion is violated the phase transition will generically remain sharp, but
the critical behavior will be different from the clean case. There are two possible scenarios, a
finite-randomness critical point or an infinite-randomness critical point. A critical point is of
finite-randomness type if, under coarse graining, the system stays disordered on all length scales
with the effective strength of the randomness approaching a finite constant. The probability
distributions of thermodynamic observables reach a finite width in the thermodynamic limit, i.e.,
they are not self-averaging.\cite{Aharony:96,Wiseman:98} From a renormalization group point of
view this means there is a critical fixed point with finite disorder strength. At a
finite-randomness critical point, the thermodynamic observables obey standard power-law scaling
behavior, but with exponents different from the exponents of the corresponding clean system. The
other scenario, an infinite-randomness critical point, occurs if the effective disorder strength
in the system grows without limit under coarse graining. The system looks more and more disordered
on larger and larger length scales, i.e., it is described by a renormalization group fixed point
with infinite disorder. The probability distributions of the thermodynamic observables become very
broad (even on the logarithmic scale) and their widths diverge when approaching the critical
point. The scaling behavior is of activated rather than of conventional power-law type. A famous
example of an infinite-randomness critical point occurs in the McCoy-Wu
model,\cite{McCoy:68,McCoy:69a} a $2d$ Ising model with bond disorder perfectly correlated in one
dimension and uncorrelated in the other. Recently, infinite-randomness critical points have also
been found in several $1d$ random quantum spin chains and two-dimensional random quantum Ising
models.\cite{Ma:79,Bhatt:82,Fisher:92,Fisher:94,Fisher:95,Young:96,Pich:98,Motrunich:00}

Disorder does not only influence the physics at the critical point itself, but also produces
interesting effects close to it. These effects are known as Griffiths phenomena, a topic that has
regained considerable attention in recent years. Griffiths phenomena are non-perturbative effects
produced by rare disorder fluctuations close to a phase transition. They can be understood as
follows: Generically, the critical temperature $T_c$ of a disordered system is lower than its
clean value, $T_c^0$. In the temperature interval $T_c<T<T_c^0$, the bulk system is in the
disordered phase. On the other hand, in an infinite size sample, there is an exponentially small,
but finite probability for finding an arbitrary large region devoid of impurities. Such a region,
a 'Griffiths island', can develop local order while the bulk system is still disordered. Due to
its size, such an island will have very slow dynamics because flipping it requires changing of the
order parameter over a large volume, which is a slow process. Griffiths\cite{Griffiths:69} showed
that the presence of the locally ordered islands produces an essential
singularity\cite{Griffiths:69,Bray:89} in the free energy in the whole region $T_c<T<T_c^0$, which
is now known as the Griffiths region or the Griffiths phase.\cite{Randeria:85} In generic
classical systems the Griffiths singularity is weak, and it does not significantly contribute to
the \emph{thermodynamic} observables. In contrast, the long-time dynamics is dominated by these
rare regions. Inside the Griffiths phase the spin autocorrelation function $C(t)$ decays as $\ln
C(t)\sim -(\ln t)^{d/(d-1)}$ for Ising systems\cite{Randeria:85,Dhar,Dhar:88,Bray:88a,Bray:88b}
and as $\ln C(t)\sim -t^{1/2}$ for Heisenberg systems.\cite{Bray:88a,Bray:87} These results were
recently confirmed by more rigorous calculation for the equilibrium\cite{Dreyfus:95,Gielis:95} and
dynamic\cite{Cesi:97a,Cesi:97b} properties of disordered Ising systems.

There are numerous systems where the disorder is not point like, but is realized through, e.g.,
dislocations or grain boundaries. This extended disorder in a $d$-dimensional system can often be
modeled by defects perfectly correlated in $d_C$ dimensions and uncorrelated in the remaining
$d_\perp=d-d_C$ dimensions.  It is generally agreed that extended disorder will have even stronger
effects on a phase transition than  point-like impurities. Nevertheless, the fate of the
transition in the presence of the extended impurities is not settled. Early renormalization group
analysis\cite{Lubensky:75} based on a single expansion in $\epsilon=4-d$ did not produce a
critical fixed point, leading to the conclusion that the phase transition is either smeared or of
first order.\cite{Rudnick:78,Andelman:85} Later
work\cite{Dorogovtsev:80,Boyanovsky:82,DeCesare:94} which included an expansion in the number of
correlated dimensions $d_C$ lead to a fixed point with conventional power law scaling. Subsequent
Monte-Carlo simulations of a $3d$ Ising model with planar defects provided further support for a
sharp phase transition scenario.\cite{Lee:92} Notice, however, that the perturbative
renormalization group calculations missed all effects coming form the rare regions. These effects
were extensively studied for the above-mentioned McCoy-Wu model. While it was believed for a long
time that the phase transition in this model is smeared, it was later found to be sharp, but of
infinite-randomness type. \cite{McCoy:69b,Fisher:92, Fisher:95}  Based on these findings, there
was a general belief that a phase transition will remain sharp even in the presence of extended
disorder.

Recently, it has been shown that this belief is not true. A theory\cite{Vojta:03a,Vojta:03b} based
on extremal statistics arguments has predicted that impurities correlated in a sufficiently high
number of dimensions will generically smear the phase transition. The predictions of this theory
were confirmed in simulations of mean-field type models\cite{Vojta:03a, Vojta:03b} but up to now,
a demonstration of the smearing in a more realistic short-range model has been missing.

In this paper, we therefore present results of large-scale Monte-Carlo simulations for a $3d$
Ising model with planar defects and nearest-neighbor interactions in both the correlated and
uncorrelated dimensions. These simulations show that the sharp phase transition is indeed
destroyed by the extended disorder. The smearing of the transition is a consequence of a mechanism
similar to but stronger than the Griffiths phenomena. In an Ising system with planar defects true
static long-range order can develop on rare islands devoid of impurities. As a consequence, the
order parameter becomes spatially very inhomogeneous and its average develops an exponential
dependence on temperature. This paper is organized as follows. In section \ref{sec:Model}, the
model is introduced and the mechanism of the smearing is explained. Section \ref{sec:NR} is
devoted to the results of the Monte-Carlo simulations and a comparison with the theoretical
predictions. In Section \ref{sec:Conc}, we present our conclusions and discuss a number of open
questions.

\section{The Model}
\label{sec:Model}

\subsection{3D Ising model with planar defects}
\label{sec:3DIsing} Our starting point is a $3d$ Ising model with planar defects. Classical Ising
spins $S_{ijk}=\pm 1$ reside on a cubic lattice. They interact via nearest-neighbor interactions.
In the clean system all interactions are identical and have the value $J$.  The defects are
modeled via 'weak' bonds randomly distributed in one dimension (uncorrelated direction). The bonds
in the remaining two dimensions (correlated directions) remain equal to $J$. The system
effectively consists of blocks separated by parallel planes of weak bonds. Thus, $d_\bot=1$ and
$d_C=2$. The Hamiltonian of the system is given by:
\begin{eqnarray}
\label{eq:1} H=&&-\sum_{{i=1,\dots,L_\bot} \atop
{j,k=1,\dots,L_C}}J_iS_{i,j,k}S_{i+1,j,k}\nonumber\\ && - \sum_{{i=1,\dots,L_\bot} \atop
{j,k=1,\dots,L_C}}J(S_{i,j,k}S_{i,j+1,k}+S_{i,j,k}S_{i,j,k+1}),
\end{eqnarray}
where $L_\bot$($L_C$) is the length in the uncorrelated (correlated) direction, $i$, $j$ and $k$
are integers counting the sites of the cubic lattice, $J$ is the coupling constant in the
correlated directions and $J_i$ is the random coupling constant in the uncorrelated direction. The
$J_i$ are drawn from a binary distribution:
\begin{equation}
\label{eq:2} J_i=\left\{
\begin{array}{clc}
 cJ & \textrm{with probability} & p\\ J  & \textrm{with probability} & 1-p
\end{array}
\right .
\end{equation}
characterized by the concentration $p$ and the relative strength $c$ of the weak bonds ($0<c\le
1$). The fact that one can independently vary concentration and strength of the defects in an easy
way is the main advantage of this binary disorder distribution. However, it also has unwanted
consequences, viz. log-periodic oscillations of many observables as functions of the distance from
the critical point.\cite{Karevski:96} These oscillations are special to the binary distribution
and unrelated to the smearing considered here; we will not discuss them further. The order
parameter of the magnetic phase transition is the total magnetization:
\begin{equation}
\label{eq:3} m=\frac{1}{V} \sum_{i,j,k} \langle S_{i,j,k} \rangle,
\end{equation}
where $V=L_\bot L_C^2$ is the volume of the system, and $\langle\cdot\rangle$ is the thermodynamic
average.

Now we consider the effects of rare disorder fluctuations in the system. Similarly to the
Griffiths phenomena, there is a small but finite probability to find a large spatial region
containing only strong bonds  in the uncorrelated direction. Such a rare region can locally be
in the ordered state while the bulk system is still in the disordered (paramagnetic) phase. The
ferromagnetic order on the largest rare regions starts to emerge right below the clean critical
temperature $T_c^0$. Since the defects in the system are planar, these rare regions are
infinite in the two correlated dimensions but finite in the uncorrelated direction. This makes
a crucial difference compared to systems with uncorrelated disorder, where rare regions are of
finite extension. In our system, each rare region is equivalent to a two dimensional Ising
system that can undergo a real phase transition independently of the rest of the system. Thus,
each rare region can independently develop true static order with a non-zero static value of
the local magnetization. Once the static order has developed, the magnetizations of different
rare regions can be aligned by an infinitesimally small interaction or external field. The
resulting phase transition will thus be markedly different from a conventional continuous phase
transition. At a conventional transition, a non-zero order parameter develops as a collective
effect of the entire system which is signified by a diverging correlation length of the order
parameter fluctuations at the critical point. In contrast, in a system with planar defects,
different parts of the system (in the uncorrelated direction) will order independently, at
different temperatures. Therefore the global order will develop inhomogeneously and the
correlation length in the uncorrelated direction will remain finite at all temperatures. This
defines a smeared transition. Thus we conclude that planar defects destroy a sharp phase
transition and lead to its smearing.

\subsection{Results of extremal statistics theory}
\label{sec:Lifshitz}

In this subsection we briefly summarize the results of the extremal statistics
theory\cite{Vojta:03b} for the behavior in the 'tail' of the smeared transition, i.e., in the
parameter region where a few rare regions have developed static order but their density is
still sufficiently low so they can be considered as independent. The approach is very similar
to that of Lifshitz \cite{Lifshitz:64} and others developed for the description of the tails in
the electronic density of states. The extremal statistics theory\cite{Vojta:03b} correctly
describes the leading (exponential) behavior of the magnetization and other observables. A
calculation of pre-exponential factors would be much more complicated because one would have to
include, among other things, details of the geometry of the rare regions, surface critical
behavior \cite{surface_binder,surface_diehl} at the surfaces of the rare regions, and
corrections to finite-size scaling. This is beyond the scope of the present paper.

The probability $w$ to find a large region of linear size $L_\bot$ containing only strong bonds
is, up to pre-exponential factors:
\begin{equation}
\label{eq:4} w\sim (1-p)^{L_\bot} = e^{\log(1-p)L_\bot}.
\end{equation}
As discussed in subsection \ref{sec:3DIsing}, such a rare region develops static long-range
(ferromagnetic) order at some reduced temperature $T_c(L_\bot)$ below the clean critical reduced
temperature $T_c^0$. The value of $T_c(L_\bot)$ varies with the length of the rare region; the
longest islands will develop long-rage order closest to the clean critical point. A rare region is
equivalent to a slab of the clean system, we can thus use finite size scaling to obtain:
\begin{equation}
\label{eq:5} T_c^0-T_c(L)=|t_c(L)|=AL^{-\phi},
\end{equation}
where $\phi$ is the finite-size scaling shift exponent of the clean system and A is the amplitude
for the crossover from three dimensions to a slab geometry infinite in two (correlated) dimension
but with finite length in the third (uncorrelated) direction. The reduced temperature $t = T -
T_c^0$ measures the distance form the \emph{clean} critical point. Since the clean 3$d$ Ising
model is below its upper critical dimension ($d_c^{+}=4$), hyperscaling is valid and the
finite-size shift exponent $\phi=1/\nu$. Combining (\ref{eq:4}) and (\ref{eq:5}) we get the
probability for finding an island of length $L_\bot$ which becomes critical at some $t_c$ as:
\begin{equation}
\label{eq:6}
\begin{array}{cc}
w(t_c)\sim e^{-B|t_c|^{-\nu}} & \textrm{(for $t_c\to0-$)}
\end{array}
\end{equation}
with the constant $B=-\log(1-p)A^{\nu}$. The total (average) magnetization $m$ at some reduced
temperature $t$ is obtained by integrating over all rare regions which have $t_c>t$. Since the
functional dependence on $t$ of the local magnetization on the island is of power-law type it does
not enter the leading exponentials but only pre-exponential factors, so:
\begin{equation}
\label{eq:7}
\begin{array}{cc}
m(t)\sim e^{-B|t|^{-\nu}} & \textrm{(for $t\to0-$)}.
\end{array}
\end{equation}

Now we turn our attention to the homogeneous magnetic susceptibility. It contains two
contributions, one coming from the islands on the verge of ordering and one from the bulk system
still deep in the disordered phase. The bulk system provides a finite, non-critical background
susceptibility throughout the whole tail region of the smeared transition. In order to estimate
the second part of the susceptibility, i.e., the part coming from the islands consider the onset
of local magnetization at the clean critical point. Using eq.\ (\ref{eq:6}) for the density of
islands we can estimate:
\begin{equation}
\label{eq:8}
\begin{array}{cc}
\chi\sim\int_0^{\Lambda}dt t^{-\gamma}e^{-Bt^{-\nu}} & \textrm{(for $t\to0-$)}.
\end{array}
\end{equation}
The last integral is finite because the exponentially decreasing island density overcomes the
power-law divergence of the susceptibility of an individual island. Here $\gamma$ is the clean
susceptibility exponent and $\Lambda$ is related to a lower cutoff for the island size. Once
the first island is ordered it produces an effective background magnetic field which cuts off
any possible divergence in $\chi$. Therefore, we conclude that the homogeneous magnetic
susceptibility does not diverge anywhere in the tail of the smeared transition. However, there
is an essential singularity at the clean critical temperature produced by the vanishing density
of ordered islands. Because if this singularity one might be tempted to call this temperature
the transition temperature of our system, but this is not appropriate because at this
temperature only an infinitesimally small part of the system starts to develop a finite
magnetization while most of the system remains solidly in the nonmagnetic phase. We rather view
the clean critical temperature as the onset of the smearing region in our
model.\cite{Vojta:03b}

The spatial distribution of the magnetization in the tail region of the smeared transition is
very inhomogeneous. On the already ordered islands, the local (layer) magnetization
$m_i=(1/L_C^2)\sum_{j,k}\langle S_{i,j,k}\rangle$  is comparable to the magnetization of the
clean system. On the other hand, far away from the ordered islands $m_i$ decays exponentially
with the distance from the closest one. The probability distribution of the logarithm of the
magnetization $P[\log m_i]$ will therefore be very broad, ranging from $\log m_i = O(1)$ on the
largest islands to $\log m_t\to-\infty$ on sites very far away from any ordered islands. The
typical magnetization $m_{typ}$ can be estimated from the typical distance of a point from the
nearest ordered island. Using eq.\ (\ref{eq:6}) we get:
\begin{equation}
\label{eq:9} x_{typ}\sim e^{B|t|^{-\nu}}.
\end{equation}
At the distance $x_{typ}$ from an ordered island, the local magnetization has decayed to
\begin{equation}
\label{eq:10} m_{typ}\sim e^{-x_{typ}/\xi_0}\sim e^{-Ce^{B|t|^{-\nu}}}
\end{equation}
where $\xi_0$ is the bulk correlation length, which is finite and changes slowly throughout the
tail region of the smeared transition, and $C$ is a constant. A comparison with eq.\ (\ref{eq:7})
gives the relation between $m_{typ}$ and the thermodynamic order parameter (magnetization) $m$ as:
\begin{equation}
\label{eq:11} |\log m_{typ}|\sim \frac{1}{m}.
\end{equation}
Thus, $m_{typ}$ decays exponentially with $m$ indicating an extremely broad order parameter
distribution. In order to determine the functional form of the local order parameter distribution,
first consider a situation with just a single ordered island at the origin of the coordinate
system. For large distances $x$, the local magnetization falls off exponentially as $m(x) = m_0~
e^{-x/\xi_0}$. The probability distribution of $y=\log[m(x)]=\log m_0-x/\xi_0$ can be calculated
from
\begin{equation}
\label{eq:12} P(|y|) = \left |\frac {dN}{dy} \right| = \frac{dN}{dx} \left | \frac {dx}{dy}\right
| =\xi_0 \frac{dN}{dx}\sim \xi_0
\end{equation}
where $dN$ is the number of sites at a distance from the origin between $x$ and $x+dx$ or,
equivalently, having a logarithm of the local magnetization between $y$ and $y+dy$. Therefore, for
large distances, the probability distribution of $\log m(x)$ generated by a single ordered island
takes the form
\begin{equation}
\label{eq:13}
\begin{array}{cc}
P[\log(m)] = const. & (\textrm{for } m(x) \ll 1)~.
\end{array}
\end{equation}
In the tail region of the smeared transition our system consists of a few ordered islands whose
distance is large compared to $\xi_0$. The probability distribution of the local magnetization,
$\log(m_i)$, thus takes the form (\ref{eq:13}) with a lower cutoff corresponding to the typical
island-island distance and an upper cutoff corresponding to a distance $\xi_0$ from an ordered
island.

\subsection{Finite-size effects}
\label{sec:finite}

It is important to distinguish effects of a finite size $L_C$ in the correlated directions and a
finite size $L_\bot$ in the uncorrelated directions. If $L_\bot$ is finite but $L_C$ is infinite
static order on the rare regions can still develop. In this case, the sample contains only a
finite number of islands of a certain size. As long as the number of relevant islands is large,
finite size-effects are small and governed by the central limit theorem. However, for $t\to 0-$
very large and rare islands are responsible for the order parameter. The number $N$ of islands
which order at $t$ behaves like $N \sim L_\bot w(t)$. When $N$ becomes of order one, strong
sample-to-sample fluctuations arise. Using (\ref{eq:6}) for $w(t)$ we find that strong sample to
sample fluctuations start at
\begin{eqnarray}
\label{eq:14} |t_L| \sim \left( \frac {1}{B} \log (L_\bot) \right)^{-1/\nu}~.
\end{eqnarray}
Thus, finite size effects are suppressed only logarithmically.

Analogously, one can study the onset of static order in a sample of finite size $L_\bot$ (i.e.,
the ordering temperature of the largest rare region in this sample). For small sample size
$L_\bot$, the probability distribution $P(T_s)$ of the sample ordering temperatures $T_s$ will be
broad because some samples do not contain any large islands. With increasing sample size the
distribution becomes narrower and moves toward the clean $T_c^0$ because more samples contain
large islands. The maximum $T_s$ coincides with $T_c^0$ corresponding to a sample without
impurities. The lower cutoff corresponds to an island size so small that essentially every sample
contains at least one of them. Consequently, the width of the distribution of critical
temperatures in finite-size samples is governed by the same relation as the onset of the
fluctuations,
\begin{eqnarray}
\label{eq:15} \Delta T_{s} \sim \left ( \frac {1}{B} \log (L_\bot) \right)^{-1/\nu }~.
\end{eqnarray}

For the system under study in this paper, a finite size in the correlated direction has far less
interesting consequences. In this case the rare regions are finite in all directions and cannot
develop true static order. Therefore, the phase transition is rounded by conventional finite-size
effects in addition to the disorder induced smearing discussed in this paper.

\section{Numerical Results}
\label{sec:NR}

\subsection{The method}
\label{sec:method}

We now turn to the main part of the paper, Monte-Carlo simulations of a $3d$ Ising model with
planar bond defects  and short range interactions, as given in eq. (\ref{eq:1}). The simulations
are performed using the Wolff cluster algorithm.\cite{Wolff:89}

As discussed above, the smearing of the transition is a result of exponentially rare events.
Therefore sufficiently large system sizes are required in order to observe it. We have
simulated system sizes ranging from $L_\bot=50$ to $L_\bot=200$ in the uncorrelated direction
and from $L_C=50$ to $L_C=400$ in the remaining two correlated directions, with the largest
system simulated having a total of 32 million spins. We have chosen $J=1$ and $c=0.1$ in the
eq.\ (\ref{eq:2}), i.e., the strength of a 'weak' bond is 10\% of the strength of a strong
bond. The simulations have been performed for various disorder concentrations $p=\{0.2, 0.25,
0.3\}$. The values for concentration $p$ and strength $c$ of the weak bonds have been chosen in
order to observe the desired behavior over a sufficiently broad interval of temperatures. This
issue will be discussed in more detail in Section \ref{sec:Conc}.  The temperature range has
been $T=4.325$ to $T=4.525$, close to the critical temperature of the clean $3d$ Ising model
$T_c^0=4.511$.

Monte-Carlo simulations of disordered systems require a huge computational
effort.\cite{selke_review} For optimal performance one must thus carefully choose the number
$N_S$ of disorder realizations (i.e., samples) and the number $N_I$ of measurements during the
simulation of each sample. Assuming full statistical independence between different
measurements (quite possible with a cluster update), the variance $\sigma_T^2$ of the final
result (thermodynamically and disorder averaged) for a particular observable is given
by\cite{Ballesteros_1998a,Ballesteros_1998b}
\begin{equation}
\sigma_T^2 = (\sigma_S^2 + \sigma_I^2/N_I)/N_S
\end{equation}
where $\sigma_S$ is the disorder-induced variance between samples and $\sigma_I$ is the
variance of measurements within each sample. Since the computational effort is roughly
proportional to $N_I N_S$ (neglecting equilibration for the moment), it is then clear that the
optimum value of $N_I$ is very small. One might even be tempted to measure only once per
sample. On the other hand, with too short measurement runs most computer time would be spent on
equilibration.

In order to balance these requirements we have used a large number $N_S$ of disorder
realizations, ranging from 30 to 780, depending on the system size and rather short runs of 100
Monte-Carlo sweeps, with measurements taken after every sweep. (A sweep is defined by a number
of cluster flips so that the total number of flipped spins is equal to the number of sites,
i.e., on the average each spin is flipped once per sweep.) The length of the equilibration
period for each sample is also 100 Monte-Carlo sweeps. The actual equilibration times have
typically been of the order of $10$-$20$ sweeps at maximum. Thus, an equilibration period of
100 sweeps is more than sufficient.

\subsection{Total magnetization and susceptibility}
\label{sec:global_mag}

\begin{figure}
\includegraphics[width=\columnwidth]{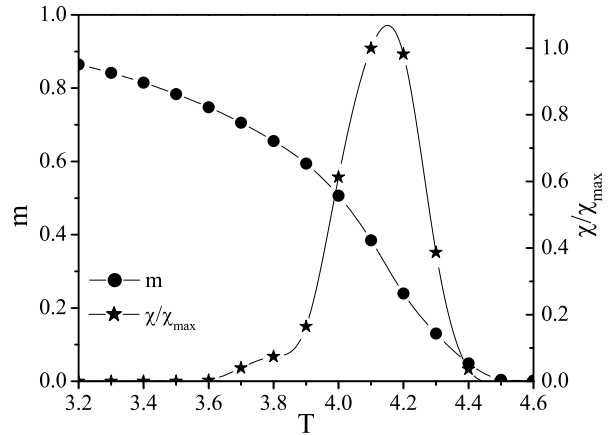}
\caption{Average magnetization $m$ and susceptibility $\chi$ (spline fit) as functions of $T$
for $L_\bot=100$, $L_C=200$ and $p=0.2$ averaged over 200 disorder realizations.}
\label{fig:mag}
\end{figure}

In this subsection we present numerical results for the total magnetization $m$  (as usual, our
Monte-Carlo estimator of $m$ is the average of the {\em absolute value} of the magnetization in
each measurement) and the homogeneous susceptibility $\chi=\partial m/\partial h$. Fig.
\ref{fig:mag} gives an overview of total magnetization and susceptibility as functions of
temperature averaged over 200 samples of size $L_\bot=100$ and $L_C=200$ with an impurity
concentration $p=0.2$.
We note that at the first glance the transition looks like a sharp phase transition with a
critical temperature between $T=4.3$ and $T=4.4$, rounded by conventional finite size effects.
In order to distinguish this conventional scenario from the disorder induced smearing of
section \ref{sec:Model}, we have performed a detailed analysis of the system in a temperature
range in the immediate vicinity of the clean critical temperature $T_c^0=4.511$.

\begin{figure}
\includegraphics[width=\columnwidth]{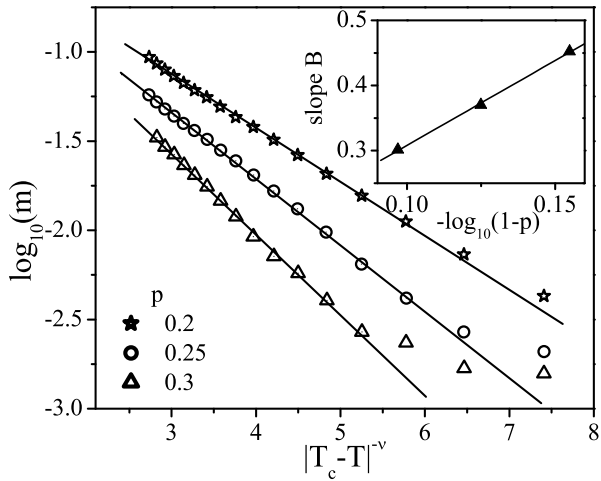}
\caption{Logarithm of the total magnetization $m$ as a function of $|T_c^0-T|^{-\nu}$
($\nu=0.627$) for several impurity concentrations $p=0.2, 0.25, 0.3$, averaged over 240
disorder realizations. System size $L_\bot=200$, $L_C=280$. The statistical errors are smaller
than a symbol size for all $\log_{10}(m)>-2.5$. Inset: Decay slope $B$ as a function of
$-\log(1-p)$.} \label{fig:logmag1}
\end{figure}

In Fig. \ref{fig:logmag1}, we plot the logarithm of the total magnetization vs.
$|T_c^0-T|^{-\nu}$ averaged over 240 samples for system size $L=200$, $L_C=280$ and three
disorder concentrations $p=\{0.2,0.25,0.3\}$.
The standard deviation of the total magnetization is below $10^{-3}$. For all three
concentrations the data follow the analytical prediction, eq. (\ref{eq:7}), over more than an
order of magnitude in $m$ with the exponent for the clean Ising model $\nu=0.627$. The
deviation from the straight line for small $m$ is due to the conventional finite size effects
(see discussion in subsection \ref{sec:finite_size}). In the inset we show that the decay
constant $B$ depends linearly on $-\log(1-p)$. This is the behavior expected from eq.
(\ref{eq:4}).

\subsection{Finite size effects and sample-to-sample fluctuations}
\label{sec:finite_size}

As discussed in subsection \ref{sec:finite} one should distinguish between two different finite
size effects, i.e., effects coming form the finite size $L_C$ in correlated direction and effects
produced by the finite size $L_\bot$ in uncorrelated direction.

\begin{figure}
\includegraphics[width=\columnwidth]{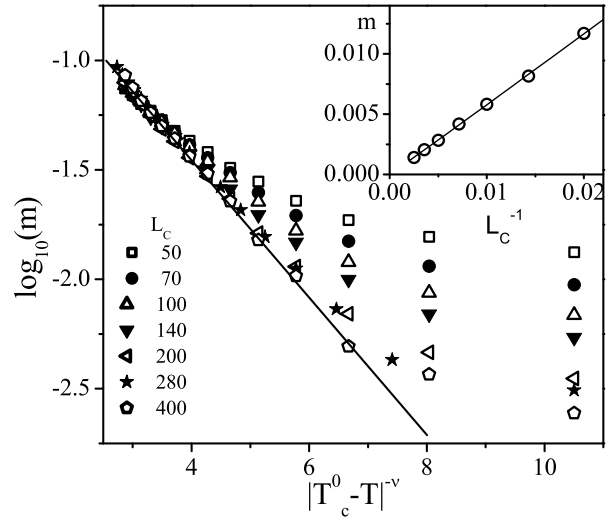}
\caption{Logarithm of the total magnetization $m$ as a function of $|T_c^0-T|^{-\nu}$
($\nu=0.627$) for disorder concentration $p=0.2$ and system sizes $L_\bot=200$, $L_C=50 \ldots
400$. The statistical errors are smaller than about a symbol size. The solid line shows the
analytic prediction, eq. (\ref{eq:7}). Inset: Total magnetization $m$ as a function of inverse
length in the correlated direction $L_C$ for $T=4.5$ ($|T-T_c^0|^{-\nu}=16.91$).}
\label{fig:logmag2}
\end{figure}

We start with analysis of the finite size effects in correlated directions, i.e. $L_C$ finite and
$L_\bot\to\infty$. The true static order on the rare regions is destroyed by the finite length of
the island in the correlated direction. For our model $d_\perp=1$ so no true static long range
order can develop. The value of $m$ measured in the simulations is thus due to fluctuations which
are governed by the central limit theorem, i.e., $m\sim V^{-1/2}$, where $V=L_\bot L_C^{2}$ is the
volume of the system. This produces a conventional finite-size rounding responsible for the
deviations of $m$ from the exponential law in Fig.\ \ref{fig:logmag1}. In Fig.\ \ref{fig:logmag2},
we investigate this finite-size effect in more detail.
This figure shows the total magnetization $m$ as a function of $|T_c^0-T|^{-\nu}$ for systems
with fixed size in the uncorrelated direction $L_\bot=200$ and various lengths in the
uncorrelated direction, $L_C=50, 70, 100, 140, 200, 280, 400$. The magnetization is averaged
over 30 to 240 disorder realizations. As expected, for high temperatures, the total
magnetization shows a strong dependence on $L_C$. The smallest systems follow the exponential
behavior (\ref{eq:7}) only over a narrow range of temperatures and then cross over to the
fluctuation determined value. If $L_C$ is increased the crossover between the exponential
behavior (\ref{eq:7}) and the fluctuation background shifts to higher temperatures. In order to
show that the fluctuation-determined value of the total magnetization $m$ at high temperatures
indeed follows the predictions of the central limit theorem, i.e. $m\sim V^{-1/2}=(L_\bot
L_C^2)^{-1/2}\sim 1/L_C$ ($L_\bot$ is constant) we plot $m$ as a function of $1/L_C$ ($T=4.5$,
$|T-T_c^0|^{-\nu}=16.91$). The numerical data shown in the inset of Fig.\ \ref{fig:logmag2} can
indeed be well fitted with a straight line. These results show that the small-$m$ deviations
from the predicted behavior (\ref{eq:7}) are indeed the result of conventional finite-size
rounding.

\begin{figure}[tb]
\includegraphics[width=\columnwidth]{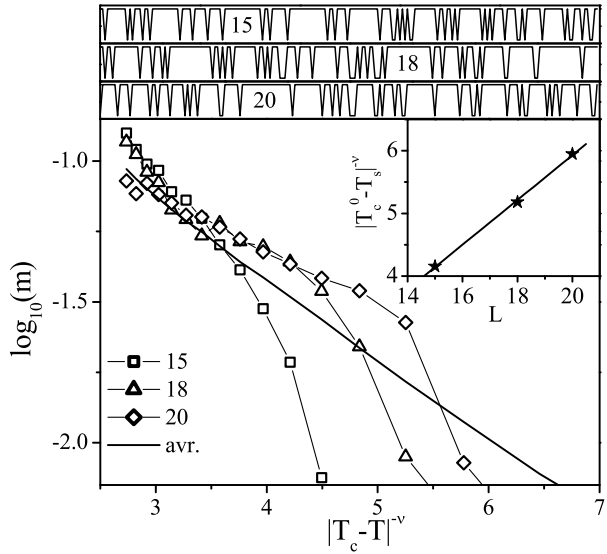}
\caption{Logarithm of the total magnetization $m$ as a function of $|T_c^0-T|^{-\nu}$ for
$L_\bot=200$, $L_C=280$ and $p=0.2$ for three different disorder realizations. The
thermodynamic statistical error of $\log_{10} m$ of a {\em single} realization is about 0.1.
Straight line represents the average over 240 disorder realizations. Upper panel: The coupling
constant $J_i$ in the uncorrelated direction as a function of $i$ for the corresponding three
disorder realizations. Numbers indicate length of the longest island $L_i$ in the uncorrelated
direction. Inset: Relation between the sample critical temperature $T_s$ and the size of the
island length, plotted as $|T_c^0-T_s|^{-\nu}$ as a function of island length.}
\label{fig:confmag}
\end{figure}

We now turn our attention to the more interesting finite size effects produced by the finite
sample length $L_\bot$ in the uncorrelated direction. For sufficiently small $L_\bot$ one
expects strong sample to sample fluctuations, as discussed in subsection \ref{sec:finite}. In
Fig. \ref{fig:confmag} we show the logarithm of the total magnetization $m$ as a function of
$|T_c^0-T|^{-\nu}$ for three typical disorder realizations.
For comparison, the upper panel of the Fig.\ \ref{fig:confmag} shows the coupling constant
$J_i$ as a function of the position $i$ for the three samples. The numbers in the graph
indicate the lengths of the longest islands $L_i$. The system size is $L_\bot=200$, $L_C=280$
with disorder concentration $p=0.2$. The solid line is the average magnetization over 240
disorder realizations. We see that all three curves qualitatively follow the average at low
temperatures but start to deviate form it at higher temperatures. The temperature $T_s$ at
which the magnetization of a sample rapidly drops is associated with the ordering of the
largest island in this sample. Numerically, we determine $T_s$ as the temperature where the
sample magnetizations falls below $1/3$ of the average magnetization. This definition contains
some amount of arbitrariness which corresponds to an overall shift of all $T_s$. However, the
leading functional dependence of $T_s$ on the size $L_i$ of the longest island in the sample is
not influenced by this shift.
In order to demonstrate  this dependence we can apply finite size scaling for the clean $3d$
Ising model (islands are regions devoid of impurities) in the slab geometry, i.e. on a sample
of length $L_i$ in one dimension and essentially infinite length in other two dimensions
($L_C\gg L_i$). In the inset of Fig. \ref{fig:confmag} we plot $|T_c^0-T_s|^{-\nu}$ as a
function of $L_i$. The data show good agreement with the finite-size scaling prediction. Figure
\ref{fig:confmag} also demonstrates that, in the tail of the smeared transition (for $T \to
T_c^0$), the average (thermodynamic) magnetization is determined by rare samples with
untypically large rare regions.

\begin{figure}[tb]
\includegraphics[width=\columnwidth]{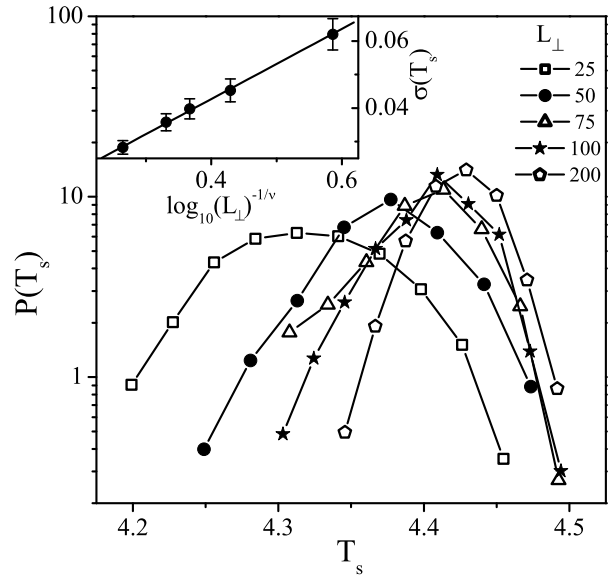}
\caption{The probability distribution of sample critical temperature $T_s$ as for different sample
lengths in the uncorrelated direction. The data shown is for system with $L_\bot=25, 50, 75, 100,
200$ and $L_C=200$. The probability distribution is calculated from 700 to 780 disorder
realizations and disorder concentration $p=0.2$. Inset: Width of the probability distribution as a
function of $\log(L_\bot)^{-1/\nu}$.} \label{fig:sdtc}
\end{figure}

In Fig. \ref{fig:sdtc}, we show the probability distribution of the sample ordering temperature
$T_s$ for system sizes $L_\bot=25, 50, 75, 100, 200$ and $L_C=200$, computed from 700 to 780
disorder realizations (the statistical error of the $T_s$ values is $\Delta T_S \lesssim
0.03$). The results are in good agreement with the predictions of subsection \ref{sec:finite},
i.e., the probability distribution of the sample critical temperature becomes narrower and
moves toward the clean critical temperature as the sample length $L_\bot$ in the uncorrelated
direction is increased. In the inset of Fig. \ref{fig:sdtc}, we show that the width of the
probability distribution (defined as its standard deviation) is proportional to
$\log(L_\bot)^{-1/\nu}$ as predicted in eq. (\ref{eq:15}).

\subsection{Local magnetization}
\label{sec:local_mag}

\begin{figure}[tb]
\includegraphics[width=\columnwidth]{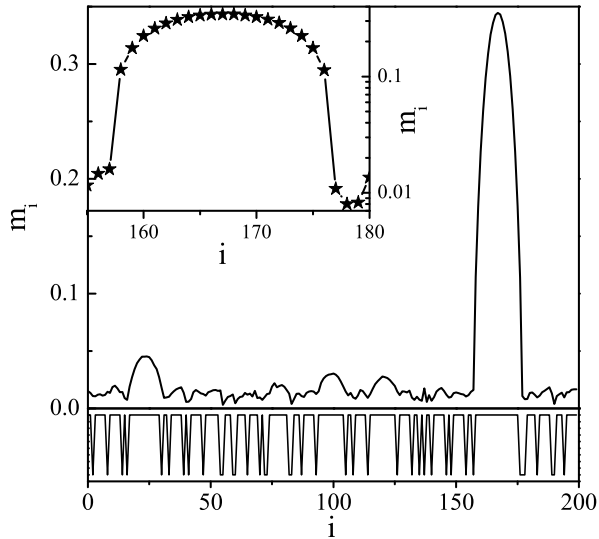}
\caption{Local magnetization $m_i$ of a particular disorder realization as a function of the
position $i$ in the uncorrelated direction (system size $L=200$, $L_C=200$ and temperature
$T=4.425$). The statistical error is approximately $5\cdot10^{-3}$. Lower panel: The coupling
constant $J_i$ in the uncorrelated direction as a function of position $i$. Inset: Log-linear
plot of the zoomed in region in the vicinity of the largest ordered island.} \label{fig:locmag}
\end{figure}

We now turn to the local (layer) magnetization $m_i$ (as for the total magnetization, our
Monte-Carlo estimator is the average of the {\em absolute values} of the layer magnetizations
for each measurement). Close to the clean critical point the system contains a few ordered
islands (rare regions devoid of impurities) typically far apart in space. The remaining bulk
system is essentially still in the disordered phase. Fig.\ \ref{fig:locmag} illustrates such a
situation.
It displays the local magnetization $m_i$ of a particular disorder realization as a function of
the position $i$ in the uncorrelated direction for the size $L_\bot=200$, $L_C=200$ at a
temperature $T=4.425$ in the tail of the smeared transition. The lower panel shows the local
coupling constant $J_i$ as a function of $i$. The figure shows that a sizable magnetization has
developed on the longest island only (around position $i=160$). One can also observe that order
starts to emerge on the next longest island located close to $i=25$. Far form these islands the
system is still in its disordered phase. In the thermodynamic limit, the local magnetization
should be exponentially small as predicted by eq. (\ref{eq:10}). However, in the simulations of a
finite size system the local magnetization has a lower cut-off which is produced by finite-size
fluctuations of the order parameter. These fluctuations are governed by the central limit theorem
and can be estimated as $m_{bulk}\approx 1/\sqrt{N_{cor}}\approx\sqrt{L^2_{cl}/L^2_C}\approx
5\cdot10^{-3}$ in agreement with the typical off-island value in Fig.\ \ref{fig:locmag}. Here,
$N_{cor}$ is the number of correlated volumes per slab as determined by the size off the Wolff
cluster. $L_{cl}$ is a typical linear size of a Wolff cluster which is, at $T=4.425$,
$L_{cl}\approx 10$). In the inset of Fig. \ref{fig:locmag} we zoom in on the region around the
largest island. The local magnetization, plotted on the logarithmic scale, exhibits a rapid
drop-off with the distance from the ordered island. This drop-off suggests  a relatively small (a
few lattice spacings) bulk correlation length $\xi_0$ in this parameter region.

As was discussed above, finite-size fluctuations of the local magnetization far form the ordered
islands mask the true asymptotic behavior for very small $m_i$. In order to verify the probability
distribution (\ref{eq:13}) of the local magnetization numerically, fluctuations have to be
suppressed sufficiently. This would require simulating very large systems whose sizes in the
correlated direction increase quadratically with the required magnetization resolution. With sizes
available in our simulations we were not able to reproduce the distribution function, eq.
(\ref{eq:13}), of $P(\log m_i)$ predicted to be constant at small $m_i$ and calculated for the
mean-field model.\cite{Vojta:03b}

\section{Conclusions}
\label{sec:Conc}

In this final section we summarize our results and discuss how the disorder induced smearing of
the phase transition found here compares to the Griffiths phenomena. We also remark on favorable
conditions for observing the disorder-induced smearing in experiments and simulations. Then we
shortly discuss differences between models with discrete and continuous symmetry. We end by
briefly addressing the question of smearing of quantum phase transitions.

We have performed large-scale Monte-Carlo simulations of a $3d$ Ising model with short-ranged,
nearest neighbor interactions and planar defects, introduced via correlated bond disorder. The
results of the simulations show that the phase transition is not sharp, but rather smeared over a
range of temperatures by the presence of the extended defects. The numerical results are in good
agreement with the theoretical predictions (see subsection \ref{sec:Lifshitz}) based on the
Lifshitz tail arguments.\cite{Vojta:03a,Vojta:03b}

The physics behind the smearing of the phase transition discussed in this paper is similar to the
physics underlying Griffiths phenomena. Both effects are produced by rare spatial regions which
are devoid of impurities and therefore locally in the ordered phase while the bulk system is still
disordered. The difference between Griffiths phenomena and disorder-induced smearing is a result
of disorder correlations. If the disorder is uncorrelated or short-range correlated, the rare
regions have finite size and cannot develop true static order. The order parameter on such a rage
region still fluctuates, albeit slowly. These slow fluctuations lead to the well known Griffiths
singularities\cite{Griffiths:69} discussed in section \ref{sec:Int}. In contrast, if the rare
regions are infinite in two or more dimensions a stronger effect arises. The rare regions can
develop true static long-range order independently of the rest of the system. The order parameter
in such a system develops very inhomogeneously, which leads to the smearing of the phase
transition. Therefore, exactly the same rare regions which would result in Griffiths phenomena if
the disorder was short-range correlated lead to the smeared phase transition in the case of
disorder correlated in two or more dimensions. In this sense the smearing of the transition takes
the place of both the phase transition and the Griffiths region. Notice that long-range
interactions increase the tendency toward smearing. If the interaction in the correlated direction
falls off as $1/r^2$ or slower, even linear defects can lead to smearing, because a $1d$ Ising
model with $1/r^2$ interaction has an ordered phase.\cite{Thouless:69,Cardy:81}

Now we turn our attention to favorable conditions for observing the smearing in numerical
simulations or experiments. This turns out to be controlled by two conditions, one for the
concentration of the impurities, and one for their strength. In order to easily observe the
smearing, the concentration of rare regions, eq. (\ref{eq:6}), has to be sufficiently large. This
requires a relatively small concentration of impurities. If the concentration of the impurities is
too high, the exponential drop-off of the island number and thus of $m$ is very steep and the
smearing effects would be very hard to observe. On the other hand, if the impurities are too weak,
the smeared transition is too close to the clean critical point and the bulk critical fluctuations
will effectively mask the smearing. Consequently, the best parameters for observing the smearing
are a small concentration of strong impurities. This has been confirmed in test calculations using
concentrations from $p=0.05$ to $0.5$. Unfavorable parameter values may also be the reason why no
smearing has been observed in previous simulations.\cite{Lee:92,Berche:98} Specifically, in Ref.
\onlinecite{Lee:92}, simulations have been performed using a high concentration $p=0.5$ of weak
impurities ($\Delta J/J=0.1$). The relatively small system sizes (up to $L=27$) in that simulation
were probably not sufficient to observe the smearing.

The next remark concerns models with continuous order parameter symmetry. As pointed out above,
the smearing of the phase transition is caused by static order on the rare regions. Thus, systems
with continuous order parameter symmetry and short-range interactions would exhibit smearing of
the phase transition only if the disorder is correlated in three or more
dimensions.\cite{Mermin:66} Again, long-range interactions increase the tendency toward smearing.
It is known\cite{Bruno:01} that classical XY and Heisenberg systems in $1d$ and $2d$ develop long
range order only if the interaction falls off more slowly than $1/r^{2d}$. Therefore a system with
linear (planar) defects would show smearing of the phase transition if the interactions in the
correlated direction would fall off more slowly then $1/r^2$ ($1/r^4$).

We end our discussion with the brief remark about smearing of quantum phase transitions in
disordered itinerant electronic systems. Each quantum phase transition can be mapped to a
classical phase transition in higher dimension, with imaginary time acting as additional
dimension. For dirty itinerant ferromagnets the effective interaction between the spin
fluctuations in the imaginary time direction falls off as $1/\tau^2$,  and the disorder is
correlated in this direction.\cite{Hertz:76} Therefore, the dirty itinerant ferromagnetic
transition is smeared even for point-like defects.\cite{Vojta:03b}

In conclusion, we have presented results of Monte-Carlo simulations of a $3d$ Ising model with
short-range interactions and planar defects. The numerical results show that the perfect disorder
correlations in two dimensions destroy the sharp magnetic phase transition leading to a smeared
transition at which the magnetization gradually develops over range of temperatures.

\begin{acknowledgments}
We acknowledge support from the University of Missouri Research Board.
\end{acknowledgments}

\bibliography{biblio}

\end{document}